\documentclass[conference]{IEEEtran}
\usepackage{cite}

\ifCLASSINFOpdf
\else
\fi
\usepackage{url}

\usepackage{booktabs}
\usepackage{tikz,pgfplots}
\usepackage{tcolorbox}

\usepackage{amsmath,amssymb,amsfonts}

\begin{document}
%
\title{Science-Software Linkage:\\ The Challenges of Traceability between Scientific Knowledge and Software Artifacts}

\author{\IEEEauthorblockN{Hideaki Hata\IEEEauthorrefmark{1},
Jin L.C. Guo\IEEEauthorrefmark{2},
Raula Gaikovina Kula\IEEEauthorrefmark{3},
Christoph Treude\IEEEauthorrefmark{4}}
\IEEEauthorblockA{\IEEEauthorrefmark{1}Shinshu University,
hata@shinshu-u.ac.jp}
\IEEEauthorblockA{\IEEEauthorrefmark{2}McGill University,
jguo@cs.mcgill.ca}
\IEEEauthorblockA{\IEEEauthorrefmark{3}Nara Institute of Science and Technology,
raula-k@is.naist.jp}
\IEEEauthorblockA{\IEEEauthorrefmark{3}University of Adelaide,
christoph.treude@adelaide.edu.au}
}


%


\maketitle

\begin{abstract}
Although computer science papers are often accompanied by software artifacts, connecting research papers to their software artifacts and vice versa is not always trivial. First of all, there is a lack of well-accepted standards for how such links should be provided. Furthermore, the provided links, if any, often become outdated: they are affected by link rot when pre-prints are removed, when repositories are migrated, or when papers and repositories evolve independently. In this paper, we summarize the state of the practice of linking research papers and associated source code, highlighting the recent efforts towards creating and maintaining such links. We also report on the results of several empirical studies focusing on the relationship between scientific papers and associated software artifacts, and we outline challenges related to traceability and opportunities for overcoming these challenges.
\end{abstract}


%
\IEEEpeerreviewmaketitle

\section{Introduction}

``Knowledge divide'' is a term used by Garousi and Rainer in their recent IEEE Software article to describe how the communities of software engineering practitioners and researchers are generally divided in terms of their mindsets and objectives~\cite{9190027}. They delineated the contrast between the gray literature published by practitioners (e.g., blog posts, white papers, and videos) and scientific papers published by researchers. The primary action-item to enable the knowledge transfer is to increase the accessibility, readability, understandability, and usefulness of scientific knowledge and artifacts.


Meanwhile, the \textit{open science} movement is increasingly shaping how the broad scientific community considers how to disseminate their work. Open science is an umbrella term that covers the concepts of open access, open data, and open-source software.
Among them, there has been remarkable progress towards open access, especially the access to research papers and articles, due to its benefit to accelerate the research advances and the existence of open-access archive repositories such as arXiv. While such progress will no doubt narrow the knowledge divide between practitioners and researchers~\cite{9190027}, it further reaffirms the need to access other scientific artifacts such as source code and poses the new challenges of creating and maintaining the traceability among scientific artifacts. 

One of the notable efforts towards scientific artifact traceability is Papers With Code\footnote{\url{https://paperswithcode.com/}}, a ``free resource for researchers and practitioners to find and follow the latest state-of-the-art ML papers and code'' with a mission to ``create a free and open resource with Machine Learning papers, code and evaluation tables.'' For each paper in its repository, the site automatically detects and links to its code, tasks, results, and methods. Since October 2020, this service is directly integrated into arXiv.\footnote{\url{https://blog.arxiv.org/2020/12/10/instant-access-to-code-for-any-arxiv-paper/}} A similar service called CatalyzeX\footnote{\url{https://www.catalyzex.com/}} provides a browser extension which ``finds and shows code implementations automatically for any machine learning, artificial intelligence, and deep learning research papers with code you come across while browsing Google, arXiv, Twitter, Scholar, GitHub, and other websites.'' The popularity of these services underlines the importance of traceability between scientific papers and software artifacts, and it highlights some of the challenges when papers and code live and evolve in separate silos.

The software engineering research community is also accelerating its commitment to open science.
In line with the open science policies of ACM SIGSOFT,\footnote{\url{https://github.com/acmsigsoft/open-science-policies}} the 43rd International Conference on Software Engineering (ICSE 2021) encourages the contributing authors to disclose their data and self-archive the pre- and post-prints in open and preserved repositories.\footnote{\url{https://conf.researchr.org/track/icse-2021/icse-2021-open-science-policies}}
The Empirical Software Engineering journal (EMSE) also encourages authors to make research materials open data and open source because they are fundamental for making empirical studies more transparent~\cite{MendezFernandez2019}. A recent study reported there is a positive trend in ICSE research track papers towards artifact availability~\cite{Heumuller2020}.
More directly, the Journal of Open Source Software (JOSS) publishes articles about research software, with links to software repositories along with the papers.\footnote{\url{https://joss.theoj.org/}}



Empirical research on scientific software has just begun.
Milewicz et al. conducted a survey of scientific software developers and reported that senior researchers did most of the contribution on their open-source software projects, with juniors also contributing on an ongoing basis~\cite{10.1109/MSR.2019.00069}.
Fan et al. analyzed the characteristics correlated with the popularity of academic AI repositories on GitHub~\cite{Fan2021}.

This emerging trend of the intersection of science and software has the potential to become a bridge to the knowledge gap between scientific communities and other communities. This paper discusses the implications and future challenges of this trend based on our empirical analyses of the linkage between science and software.

\section{Science-software Linkage}

We now report on the key findings from our work that studied the linkage between science and software: software documentation to academic papers, academic papers to software repositories, and code comments to academic papers.


\subsection{Linking Software Documentation to Academic Papers}

To study the linkage between academic papers and software repositories, we conducted an empirical analysis of GitHub repositories that make references to academic papers from their README files~\cite{2020arXiv200400199W}. We manually analyzed 377 software repositories as a statistically representative sample of approximately 20,000 README files referring to academic papers identified by pattern matching. In the 344 repositories that actually referred to academic papers, in 339 cases the article was publicly available including open access.

 \begin{figure}[t]
        \centering
        \begin{footnotesize}
        \begin{tikzpicture}
        \begin{axis}[
            align=center,
            y=0.5cm,
            x=0.05cm,
            enlarge y limits={abs=0.25cm},
            symbolic y coords={other, sensors, quantum, networks, robotics, natural language processing, other machine learning, computer vision, deep learning},
            axis line style={opacity=0},
            major tick style={draw=none},
            ytick=data,
            xmin = 0,
            xlabel = \# GitHub repositories,
            nodes near coords,
            nodes near coords align={horizontal},
            point meta=rawx
        ]
        \addplot[xbar,fill=gray,draw=none] coordinates {
            (92,deep learning)
            (83,other)
            (72,computer vision)
            (40,other machine learning)
            (29,natural language processing)
            (8,robotics)
            (7,networks)
            (6,quantum)
            (2,sensors)
        };
        \end{axis}
        \end{tikzpicture}
        \end{footnotesize}
        \caption{Domains of repositories that have references to academic papers.}
        \label{fig:rq2.1}
    \end{figure}
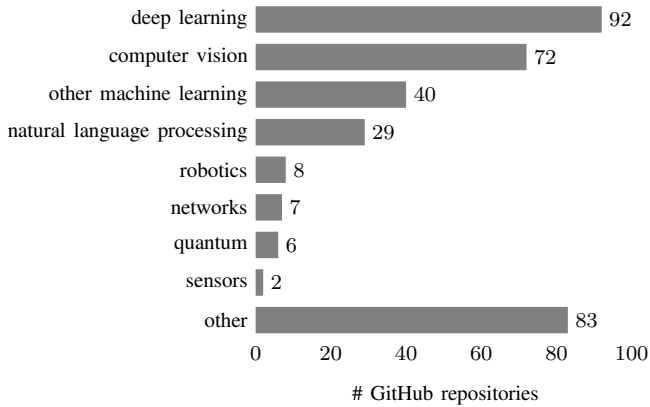

We manually coded the software domain of these 339 repositories.
As shown in Figure~\ref{fig:rq2.1}, 
we see that the most common software domain of the GitHub repositories that reference academic papers is ``deep learning'', followed by ``computer vision'' and ``other machine learning''.
In other words, machine learning is the most common, covering three quarters of the repositories in our sample.
Also, more than 20\% of the GitHub repositories referencing academic papers belong to the ``other'' code, for instance, web API, biology, or chemistry. This indicates the diversity of these GitHub repositories.

 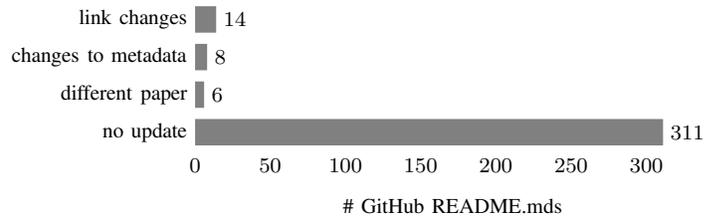
\begin{figure}[t]
        \centering
        \begin{footnotesize}
        \begin{tikzpicture}
        \begin{axis}[
            align=center,
            y=0.5cm,
            x=0.02cm,
            enlarge y limits={abs=0.25cm},
            symbolic y coords={no update, different paper, changes to metadata, link changes},
            axis line style={opacity=0},
            major tick style={draw=none},
            ytick=data,
            xlabel = \# GitHub README.mds,
            xmin = 0,
            nodes near coords,
            nodes near coords align={horizontal},
            point meta=rawx
        ]
        \addplot[xbar,fill=gray,draw=none] coordinates {
            (6,different paper)
            (8,changes to metadata)
            (14,link changes)
            (311,no update)
        };
        \end{axis}
        \end{tikzpicture}
        \end{footnotesize}
        \caption{Evoluiton in README files.}
        \label{fig:rq3.2}
    \end{figure}

For the publicly available referenced academic papers, a substantial number of papers, 53 (16\%), were found to have been updated, such as updating the version on arXiv.
On the other hand, as seen in Figure \ref{fig:rq3.2}, many README files did not have any updates regarding their references. Only 8\% had at least one change, with ``link changes'' being the most frequent (e.g., a link to a preprint version of the paper was replaced by a new link to the official website).

\begin{tcolorbox}
\textbf{Observation 1:} References to academic papers can be found in scientific repositories in many domains. In science repositories that refer to academic papers in README files, updates to academic papers are not reflected in software.
\end{tcolorbox}

\subsection{Linking Academic Papers back to Software Repositories}


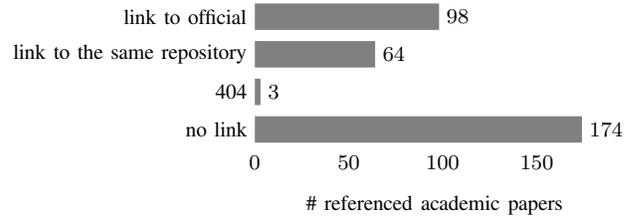
\begin{figure}[t]
        \centering
        \begin{footnotesize}
        \begin{tikzpicture}
        \begin{axis}[
            align=center,
            y=0.5cm,
            x=0.025cm,
            enlarge y limits={abs=0.25cm},
            symbolic y coords={no link,404,link to the same repository,link to official},
            axis line style={opacity=0},
            major tick style={draw=none},
            ytick=data,
            xlabel = \# referenced academic papers,
            xmin = 0,
            nodes near coords,
            nodes near coords align={horizontal},
            point meta=rawx
        ]
        \addplot[xbar,fill=gray,draw=none] coordinates {
            (98,link to official)
            (64,link to the same repository)
            (3,404)
            (174,no link)
        };
        \end{axis}
        \end{tikzpicture}
        \end{footnotesize}
        \caption{Links back from referenced academic papers.}
        \label{fig:rq2.4}
    \end{figure}

In the same study \cite{2020arXiv200400199W},
we manually investigated potential bi-directional links, i.e., links back from the referenced paper to the GitHub repository.
From the same 339 cases, we found the following patterns of links in the referenced academic papers:
the referenced paper has a link to the official repository (link to official),
the referenced paper has a link back to the GitHub repository that contains a reference to that paper (link to the same repository),
the referenced paper has a link to a software repository, but the repository cannot be accessed (404),
the referenced paper does not have a link to a software repository (no link).
As seen in Figure~\ref{fig:rq2.4}, more than half of the academic papers referenced in GitHub README files do not have a link to the GitHub repository.
Considering only the ``official'' relationship, we find that 62 out of 136 academic papers cited from official repositories refer to official repositories, while 57 papers (42\%) do not refer to official repositories. 

\begin{tcolorbox}
\textbf{Observation 2:} 
Sometimes there is no link between the academic paper and the software repository even though there is an ``official'' relationship between the paper authors and the repository owners being the same.
\end{tcolorbox}

\subsection{Linking Code Comments to Academic Papers}

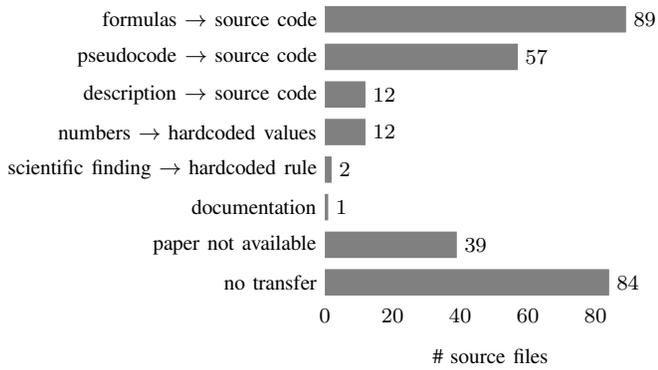
\begin{figure}
\centering
\begin{footnotesize}
\begin{tikzpicture}
\begin{axis}[
    align=center,
    y=0.5cm,
    x=0.045cm,
    enlarge y limits={abs=0.25cm},
    symbolic y coords={no transfer,paper not available,documentation,scientific finding $\rightarrow$ hardcoded rule,numbers $\rightarrow$ hardcoded values,description $\rightarrow$ source code,pseudocode $\rightarrow$ source code, formulas $\rightarrow$ source code},
    axis line style={opacity=0},
    major tick style={draw=none},
    ytick=data,
    xmin = 0,
    xlabel = \# source files,
    nodes near coords,
    nodes near coords align={horizontal},
    point meta=rawx
]
\addplot[xbar,fill=gray,draw=none] coordinates {
    (89,formulas $\rightarrow$ source code)
    (57,pseudocode $\rightarrow$ source code)
    (12,description $\rightarrow$ source code)
    (12,numbers $\rightarrow$ hardcoded values)
    (2,scientific finding $\rightarrow$ hardcoded rule)
    (1,documentation)
    (39,paper not available)
    (84,no transfer)
};
\end{axis}
\end{tikzpicture}
\end{footnotesize}
\caption{Knowledge transfer from referenced publications to source code.}
\label{fig:quali-transfer}
\end{figure}

A large-scale study of links in source code comments reported that academic papers are referenced in source code comments~\cite{10.1109/ICSE.2019.00123}.
To study the relationship between academic papers and source code files, we conducted an empirical analysis of source code comments that refer to academic papers~\cite{2019arXiv191006932I}.
We identified more than 10,000 code comments that were likely to refer to academic papers by using a named entity recognition technique to identify terms related to academic papers, such as author names, paper titles, journal names, and publication years. Of 372 comments in a statistically representative sample, we manually validated that 305 actually referred to academic papers.

Figure \ref{fig:quali-transfer} shows the result of manual coding of these 305 comments into the eight types of knowledge transfer shown below.
\begin{itemize}
    \item \textit{formulas $\rightarrow$ source code}: formulas or equations presented in the referenced publication are transformed into source code. 
    \item \textit{pseudocode $\rightarrow$ source code}: pseudocode presented in the referenced publication is transformed into source code. 
    \item \textit{description $\rightarrow$ source code}: textual descriptions presented in the referenced publication are transformed into source code. 
    \item \textit{numbers $\rightarrow$ hardcoded values}: numbers related to experimental setup (e.g., variables) or numerical findings reported in the referenced publication are added to the source code as hardcoded values.
    \item \textit{scientific finding $\rightarrow$ hardcoded rule}: scientific findings presented in the referenced publication are implemented in the source code as a hardcoded rule.
    \item \textit{documentation}: knowledge from the referenced publication is transferred for documentation only.
    \item \textit{paper not available}: we cannot determine the type of knowledge transfer since the referenced publication is not open-access.
    \item \textit{no transfer}: no knowledge from the referenced publication is transferred to the source code.
\end{itemize}
We observed that formulas or equations from the referenced scientific publications are frequently implemented in the source code, accounting for 30\% of the cases in our sample, followed by pseudocode implementation (19\%).
Some source files (28\%) with publication citations have no outside knowledge transferred to their source code since the corresponding publications provide only background or additional information associated with them.
Since a significant number (13\%) of the publications referenced in the source files are hidden behind paywalls, especially books and academic papers published in journals, we cannot investigate knowledge transfer in these cases.

\begin{table*}
    \small
    \centering
    \caption{Top 20 referenced publications}
    \label{tab:top20papers}
    \begin{tabular}{rp{22pc}rp{6pc}r}
        \toprule
        \textbf{rank} & \textbf{publication} & \multicolumn{2}{l}{\textbf{\# code citations}} & \textbf{\# paper citations} \\
        \midrule
        1 & C. Loeffler, A. Ligtenberg, G.S. Moschytz, \textbf{Practical fast 1-D DCT algorithms with 11 multiplications},  Proceedings of 1989 International Conference on Acoustics, Speech, and Signal Processing (1989) & 146 & (C 105, C++ 19, Java 1, JavaScript 21) & 1,049 \\
        1 & S. Reiter, A. Vogel, I. Heppner, M. Rupp, G. Wittum, \textbf{A massively parallel geometric multigrid solver on hierarchically distributed grids}, Computing and Visualization in Science (2013) & 146 & (C++ 146) & 49 \\
        3 & S. Tomov, J. Dongarra, \textbf{Accelerating the reduction to upper Hessenberg form through hybrid GPU-based computing}, University of Tennessee Computer Science Technical Report (2009) & 75 & (C++ 75) & 31  \\
        4 & V.I. Lebedev, D.N. Laikov, \textbf{A quadrature formula for the sphere of the 131st algebraic order of accuracy}, Doklady Mathematics (1999) & 71 & (C++ 71) & 522  \\
        5 & R. Sedgewick, \textbf{Algorithms, 2nd Edition}, Addison-Wesley (1988) & 67 & (C 61, C++ 6) & 5,335 \\
        5 & E.R. Fiala, D.H. Greene, \textbf{Data compression with finite windows}, Communications of the ACM (1989) & 67 & (C 61, C++ 6) & 300 \\
        7 & P. L'Ecuyer, \textbf{Maximally equidistributed combined Tausworthe generators}, Mathematics of Computation (1986) & 63 & (C 59, C++ 4) & 326 \\ 
        8 & M. Matsumoto, T. Nishimura, \textbf{Mersenne twister: a 623-dimensionally equidistributed uniform pseudo-random number generator}, ACM Transactions on Modeling and Computer Simulation (1998) & 57 & (C 24, C++ 14, Java 19) & 6,752 \\
        8 & P. L'Ecuyer, \textbf{Tables of maximally equidistributed combined LFSR generators}, Mathematics of Computation (1999) & 57 & (C 56, C++ 1) & 225  \\
        10 & J.P. Snyder, \textbf{Map projections--A working manual}, U.S. Geological Survey Professional Paper (1987) & 55 & (C++ 2, JavaScript 53) & 1,736 \\
        10 & P. Deutsch, \textbf{DEFLATE compressed data format specification version 1.3}, RFC 1951 (1996) & 55 & (C 51, C++ 4) & 714 \\
        12 & J.M. Robson, \textbf{Bounds for some functions concerning dynamic storage allocation}, Journal of the ACM (1974) & 51 & (C 44, C++ 7) & 84 \\
        13 & J.C.R. Bennett, H. Zhang, \textbf{Hierarchical packet fair queueing algorithms}, IEEE/ACM Transactions on Networking (1997) & 50 & (C 50) & 605 \\
        13 & I. Stoica, H. Abdel-Wahab, \textbf{Earliest eligible virtual deadline first: A flexible and accurate mechanism for proportional share resource allocation}, Technical Report (1995) & 50 & (C 50) & 60 \\
        15 & M. Matsumoto, Y. Kurita, \textbf{Twisted GFSR generators II}, ACM Transactions on Modeling and Computer Simulation (1994) & 49 & (C 47, C++ 2) & 230 \\
        16 & M. Matsumoto, Y. Kurita, \textbf{Twisted GFSR generators}, ACM Transactions on Modeling and Computer Simulation (1992) & 47 & (C 47) & 218 \\
        17 & S. Muchnick, \textbf{Advanced compiler design and implementation}, Academic Press, Morgan Kaufmann Publishers (1997) & 43 & (C 43) & 3,908 \\
        17 & R.J. Gowersk, M. Linke, J. Barnoud, T.J.E. Reddy, M.N. Melo, S.L. Seyler, J. Domanski, D.L. Dotson, S. Buchoux, I.M. Kenney, O. Beckstein, \textbf{MDAnalysis: a Python package for the rapid analysis of molecular dynamics simulations}, Proceeding of 15th Python in Science Conference (2016) & 43 & (Python 43) & 167   \\
        17 & Y. Wang, I.H. Witten, \textbf{Modeling for optimal probability prediction}, Proceedings of 19th International Conference on Machine Learning (2002) & 43 & (Java 43) & 60  \\
        17 & Y. Wang, \textbf{A new approach to fitting linear models in high dimensional spaces}, PhD Thesis, University of Waikato, NZ (2000) & 43 & (Java 43) & 48 \\
        \bottomrule
    \end{tabular}%
\end{table*}

Table~\ref{tab:top20papers} shows the top 20 referenced academic papers identified by using the same name entity recognition technique from 11,274 code comments.
We see that some academic papers that do not have a large number of citations from other papers are cited in many source code comments.

Taking a closer look at three related papers in this top 20, that is, \textit{Twisted GFSR generators} ranked in 15th and 16th places, and \textit{Mersenne twister} ranked in 8th place, we see that the Mersenne twister is extended from the previous Twisted GFSR~\cite{Matsumoto:1998:MTE:272991.272995}, and is the industry standard algorithm for generating random samples~\cite{Marsland:2014:MLA:2692349}.
Considering this fact, referencing the paper of Mersenne twister to implement the algorithm seems to be appropriate, instead of (only) referencing the older papers of Twisted GFSR.
If developers are not intentionally avoiding the Mersenne twister algorithm, there seems to be an issue of potentially obsolete knowledge. 
If these repositories update to Mersenne twister, there could be practical impact for the users of the software.

\begin{tcolorbox}
\textbf{Observation 3:} There are various types of knowledge transfer from academic papers to software modules. We identified one case where a new academic paper with an updated algorithm was not reflected in some software.
\end{tcolorbox}

\section{Traceability Challenges}
Our studies highlight several key considerations when approaching traceability between scientific knowledge and artifacts, ranging from the link creation and identification to link maintenance.

\paragraph{Link creation} 
In traditional software development, the scope, format, and purpose of trace links should be defined in the traceability strategy as part of the project planning effort. For scientific work, however, people tend to inject links in a rather ad-hoc manner. We have observed that README files of the software repositories and source code comments are the dominant places to contain links to scientific papers. Nevertheless, without a clear objective from the link creators and expectations from the link users, it is hard to guarantee that those trace links would fulfill their purpose.    

For the links from the academic paper, we have only observed links to the software repositories front page. There are hardly any fine-grained links between more detailed knowledge in the papers and the components/files in the repositories. To further support reproducibility and knowledge transfer, we believe there is a need for more precise and bi-directional links between papers and code.

\paragraph{Link identification} 
Since there is no established convention on where and how to put trace links, the variance on existing links causes considerable challenges in identifying and processing those links in an automated fashion.
References to academic papers can be found in any software documentation other than README and source code comments. For example, the preprint of the EMSE paper on empirical analysis of Git diffs~\cite{Nugroho2020} was discussed in an issue, and changes had been merged into the code base.\footnote{\url{https://github.com/gitextensions/gitextensions/issues/6991}}
Consequently, it still takes much manual effort to examine if the existing links can serve the purpose of knowledge transfer.

To reason the connections between scientific knowledge and corresponding software, we call for tools to properly manage and compare the meta-data of both papers and software, and even novel techniques to represent different artifacts and compare their relatedness. Automated traceability tools used in the context of software development can potentially support the link identification for scientific knowledge and software as well.

\paragraph{Link maintenance} 
Once created, trace links face the risk of decay. The links for science-software traceability are no exception. This challenge can also be alleviated by more rigorous link management support: whenever the source or target of a link (either the paper or the software) is updated, the link itself should be revised to reflect the updated information. 

Meanwhile, a certain extent of the impact of scientific knowledge can be reflected in its citations and code reuses. If the paper is revised or the software is updated, it is extremely difficult to propagate the changes on its knowledge transfer chain. Future investigation on the impact of asynchronous information update for science-software linkage is needed.    


\section{Conclusion}
Finding the linkage between scientific knowledge in the form of research papers and software artifacts in the form of software code is not trivial.
There exist efforts to link research papers and associated source code, such as Papers with Code and CatalyzeX.
However, several empirical studies conducted on these linkages raise challenges and research opportunities in regards to link creation, identification of linkages, and how to ensure that these linkages are kept up-to-date.



%

\end{document}